\documentclass[10pt]{iopart}
\usepackage{iopams}
\usepackage{cite}
\usepackage[dvips]{graphicx}
\begin{document}

\title[Chemical-potential standard for atomic Bose-Einstein condensates]
{Chemical-potential standard for atomic Bose-Einstein condensates}

\author{Sigmund Kohler\dag\ and Fernando Sols\ddag  }

\address{\dag\ Institut f\"ur Physik, Universit\"at Augsburg,
        Universit\"atsstr.~1, D-86135 Augsburg, Germany}
\address{\ddag Departamento de F\'\i sica Te\'orica de la Materia Condensada
        and Instituto ``Nicol\'as Cabrera'', Universidad Aut\'onoma de Madrid,
        E-28049 Madrid, Spain}

\ead{\mailto{sigmund.kohler@physik.uni-augsburg.de},
     \mailto{fernando.sols@uam.es}}

\begin{abstract}
When subject to an external time periodic perturbation of
frequency $f$, a Josephson-coupled two-state Bose-Einstein
condensate responds with a constant chemical potential difference
$\Delta \mu=khf$, where $h$ is Planck's constant and $k$ is an
integer. We propose an experimental procedure to produce ac-driven
atomic Josephson devices that may be used to define a standard of
chemical potential. We investigate how to circumvent some of the
specific problems derived from the present lack of advanced atom
circuit technology. We include the effect of dissipation due to
quasiparticles, which is essential to help the system relax
towards the exact Shapiro resonance, and set limits to the range
of values which the various physical quantities must have in order
to achieve a stable and accurate chemical potential difference
between the macroscopic condensates.
\end{abstract}
\pacs{%
03.75.Lm, 
74.50.+r, 
05.30.Jp, 
06.20.Fn. 
}
\section{Introduction}

The realization of Bose-Einstein condensation in dilute atomic
gases \cite{Anderson1995a, Davis1995a, Bradley1995a} has opened
the possibility of investigating the macroscopic behaviour of
coherent quantum matter in a new class of physical systems
different from superconductors and helium liquids. Together with
vortex formation, the Josephson effect between weakly coupled Bose
condensates is probably the most characteristic signature of
superfluidity. Initially predicted \cite{Josephson1962a} and
observed \cite{ande63,shap63} in the context of superconductivity,
the Josephson effect has also been observed in superfluid $^3$He
\cite{aven88,pere97}, a fact that underlies the fundamental nature
of quantum behaviour at the macroscopic scale. Atomic
Bose-Einstein condensates (BECs) have it in common with
superconductors and superfluids that their most remarkable
properties are derived from the existence of a macroscopic wave
function. Because of the profound analogies between these
gauge-symmetry broken systems, the possibility of observing the
Josephson effect in trapped atomic gases was early recognized
\cite{dalf96,Smerzi1997a,Zapata1998a,Sols1999a}. Although some
preliminary evidence for the existence of the Josephson effect in
atomic gases already exists \cite{Anderson1998a,Cataliotti2001a},
it is fair to say that, compared with its superconducting and
superfluid counterparts, the exploration of the physics of weakly
linked atomic condensates is still in its infancy. Given the
potential richness and the convenient tunability of BEC setups, it
is clear that decided progress must be made to investigate this
whole new class of problems. At present however, an important
limitation is that the technology of atom circuits (also called
``atomtronics") is still moderately developed, although a bright
future can already be foreseen \cite{cass00,hans01}. In this
sense, proposals that do not rely on the feasibility of controlled
atom transport are particularly welcome.

One of the most remarkable properties of superconducting devices
is the existence of Shapiro resonances. When two weakly coupled
superconductors are subject to a voltage difference that is the
sum of a dc component $V$ and a periodic signal $v \sin (2 \pi f
t)$, a continuous range of nonzero dc currents are possible if
\begin{equation} \label{resonance}
V=\frac{h}{2e}k f,
\end{equation}
where $2e$ is the Cooper pair charge, $h$ is Planck's constant,
and $k$ is an integer \cite{shap63,Barone}. A more common setup is
one in which, due to the existence of a large impedance in series,
an external current source is applied.  If the external current
consists of a dc contribution and a weak ac perturbation of
frequency $f$, the Josephson link displays a dc $V$-$I$
characteristic with voltage plateaus at values satisfying the
resonance condition~\eref{resonance}. The height of the Shapiro
steps linking the voltage plateaus provides a method to measure
the constant of nature $2e/h$ with such precision and universality
\cite{bloc70,fult73} that, since 1972, the reversed view has been
adopted whereby $2e/h$ is assumed to be known
and \Eref{resonance} is used to define a standard unit of voltage
\cite{Barone,tayl67,pope92}.

In this paper we address the question of whether it is possible to
prepare a BEC Josephson junction (BJJ) that satisfies the
resonance condition \eref{resonance} in a stable form. Another way
of phrasing the problem is whether macroscopic self-trapping of a
large population imbalance \cite{Raghavan1999b,Zapata2003a} may be
stabilized against dissipation by an oscillating perturbation. A
condensate in a time-dependent trap \cite{Gardiner1997a}, and
particularly a double condensate \cite{Salmond2002a}, has been
shown to provide a convenient test ground for quantum chaos. Here
we face a different regime which predominantly involves regular
motion with dissipation. The challenge is to identify the
conditions under which dissipation can be exploited to let the
system evolve towards one of the stable resonance islands.

At present we have a limited understanding of the dissipation
mechanisms operating in a BJJ which render its macroscopic
phase-number dynamics non-conservative. We know that contributions
may arise at least from incoherent exchange of normal atoms
\cite{Zapata1998a}, creation of quasiparticles by the fluctuating
condensate \cite{Zapata2003a}, or spontaneous atom losses
\cite{sina00,Kohler2002a}. As long as the net atom loss stays
relatively small, its effect on the phase dynamics is similar in
many respects to that of incoherent particle exchange, both
yielding Ohmic dissipation under a wide range of circumstances.

We wish to stress that the physics discussed in this paper applies
both to double-well condensates, displaying the external Josephson
effect, and to optically coupled two-component atomic condensates,
which exhibit the internal Josephson effect \cite{Sols1999a}. A
central concept in the forthcoming discussion is the application
of a time-dependent external potential. This may be achieved by
the application of suitably designed time-dependent magnetic
fields or dipole forces. Another essential ingredient, not
considered in Ref. \cite{Raghavan1999b},
is that the Josephson current can counteract the dissipative current,
thereby permitting the existence of stationary Shapiro resonances.
Moreover, by providing an effective friction to the phase dynamics,
the dissipative current ensures that the system gets arbitrarily close
to the exact resonance at sufficiently long times.

We want to remark that no fundamental reason prevents the
possibility of exploring this novel approach to Shapiro resonances
in the context of superconductors. There, capacitive couplings
might be used to control or measure voltage differences.

Section \ref{sec:formulation} is devoted to a quantitative
formulation of the problem. We know that each phase-locked
solution generates an attractive basin in phase space. We raise
the question of how dissipation can be exploited to let the system
land successfully on that region never to escape from it except
for quantum tunnelling decay.
The solution is offered in Section \ref{sec:solutions}, where we
investigate the dissipative dynamics in phase space.  Later, in
Section \ref{sec:probability}, we provide an estimate of the
probability that a system prepared with a large chemical potential
difference evolves in such a way that, in the process of decaying, it
is trapped in a classically stable well. The stochasticity comes from
the intrinsically random nature of the choice of phase made by the
double condensate system shortly after the connection
\cite{Zapata2003a}.  Section \ref{sec:quantum} is devoted to the
effect of quantum fluctuations, which may cause macroscopic tunnel
decay. The combined analysis of Sections
\ref{sec:formulation}--\ref{sec:quantum} yields a set of constraints
that must be fulfilled to achieve an optimal realization of the
predicted Shapiro resonances. A concluding discussion is provided in
Section \ref{sec:conclusion}.

\section{Formulation of the problem}
\label{sec:formulation}

We consider $N$ Bose-condensed atoms in a double-well trapping
potential $V(x,\xi)$ that depends on a harmonically time-dependent
control parameter $\xi(t)=\xi_0\cos(\Omega t)$.  In a situation
sufficiently close to equilibrium, a split condensate is formed
which can be described within a two-mode approximation.  Thus, we
use the for the field operator the ansatz $\psi(x)=\varphi_A(x) a
+\varphi_B(x)b$, where the orbital functions $\varphi_{A,B}$ are
normalized to unity.  This yields the two-mode Hamiltonian
\cite{Smerzi1997a,Sols1999a}
\begin{equation}
H=-\frac{\hbar\omega_R}{2}\left(a^\dagger b+b^\dagger a\right)
  +E_A(N_A,\xi) + E_B(N_B,\xi) .
\end{equation}
Here, $\omega_R$ denotes the effective Rabi frequency of the
two-mode problem, which is proportional to the Josephson coupling
energy \cite{Sols1999a}. It is important to note that, in the case
of optically coupled two-component BECs (internal Josephson
effect), $\omega_R$ does not generally coincide with the Rabi
frequency $\Omega_R$ governing atomic transitions. Rather, one has
$\omega_R=\Omega_R s_{AB}$, where $s_{AB}\equiv \int\rmd
x\,\varphi_A^* \varphi_B$ \cite{garc02}. It is only when the
trapping configuration and the interactions are such that
$\varphi_A(x)=\varphi_B(x)$, that the two frequencies become
identical. The Gross-Pitaevskii energy of fragment $i$,
\begin{equation}
\fl
E_i(N_i,\xi)
= N_i \!\int\!\! \rmd x\,\varphi_i^*(x)\left(-\frac{\hbar^2}{2m}\Delta +
V(x,\xi) +\frac{gN_i}{2}|\varphi_i(x)|^2\right)\varphi_i(x) ,
\end{equation}
$i=A,B$,
has inherited a time-dependence from the control-parameter $\xi(t)$.

For an analysis of the system in its classical limit, we replace
in the Heisenberg equations of motion the operators $a$, $b$ by
$\sqrt{N_A}e^{-i\phi_A}$, $\sqrt{N_B}e^{-i\phi_B}$ and expand the
Gross-Pitaevskii energies for small $\xi$ and small number
imbalance compared with the equilibrium values $N_A^0$ and $N_B^0$
(usually taken to be $N/2$). For the variables $\phi \equiv
(\phi_A-\phi_B)$ and $z \equiv (N_A-N_B)/N\equiv 2n/N$, with time
measured in units of $1/\omega_R$, this yields
\begin{eqnarray}
\dot \phi &=& \frac{z}{\sqrt{1-z^2}}\cos\phi +\Lambda z +
\varepsilon\cos(\Omega t)
    = \Delta\mu/\hbar , \label{phi:dot}
\\
\dot z &=& -\sqrt{1-z^2} \sin\phi - \gamma\dot\phi ,
    \label{z:dot}
\end{eqnarray}
with the scaled driving amplitude
\begin{equation}
\varepsilon = \frac{\xi_0}{\hbar\omega_R}\left[ \frac{\partial^2
E_A(N_A,\xi)}{\partial\xi\,\partial N_A} - \frac{\partial^2
E_B(N_B,\xi)}{\partial\xi\,\partial N_B}
\right]_{N_A=N_B=N/2,\,\xi=0}
\end{equation}
and the effective interaction constant
\begin{equation}
\Lambda = \frac{N}{2\hbar\omega_R}\left[ \frac{\partial^2
E_A(N_A,0)}{\partial N_A^2} + \frac{\partial^2
E_B(N_B,0)}{\partial N_B^2} \right]_{N_A=N_B=N/2}.
\end{equation}
Equations \eref{phi:dot}, \eref{z:dot} can also be obtained from
the classical non-rigid pendulum Hamiltonian
\cite{Smerzi1997a,Sols1999a} (hereafter energies are expressed in
units of $\hbar \omega_R$)
\begin{equation}
\label{H:pendulum} H(z,\phi,\lambda)=-\sqrt{1-z^2}\cos\phi +
\frac{1}{2}\Lambda z^2 + \varepsilon z\cos(\Omega t),
\end{equation}
where $(z,\phi)$ are canonically conjugate coordinates. For a
simplification, we have assumed a symmetric situation with
$E_A(N/2,\xi) = E_B(N/2,-\xi)$ --- the generalization is of course
straightforward and results in an additional phase drift.  The
last term in the equation of motion \eref{z:dot} has been
introduced phenomenologically to describe a dissipative current,
i.e.\ an incoherent exchange of atoms.  For high temperatures,
$k_BT\gg\Delta\mu$, this current is Ohmic, i.e.\ proportional to
the chemical potential difference, $\dot n = -G\,\Delta\mu$
\cite{Zapata1998a}.  Following the reasoning by Josephson
\cite{Josephson1962a}, the chemical potential difference is given
by the time derivative of the relative phase,
$\dot\phi=\Delta\mu/\hbar$, and thus we obtain the dissipative
term in equation~\eref{z:dot} with $\gamma=2\hbar G/N$
\footnote{At first sight there seems to be an ambiguity as to
whether dissipation should come as $-\gamma \dot{\phi}$ or,
rather, $-\gamma'z$. That the former is the correct approach can
also be inferred from a careful study of quantum dissipation
models \cite{weiss93}.  In the present context, however, the
choice is without practical consequences.}.  We will also analyze
the more general case in which $\dot n$ may not be a linear
function of $\dot \phi$, a situation which is likely to appear at
low temperatures.

\begin{figure}[t]
\centerline{\includegraphics{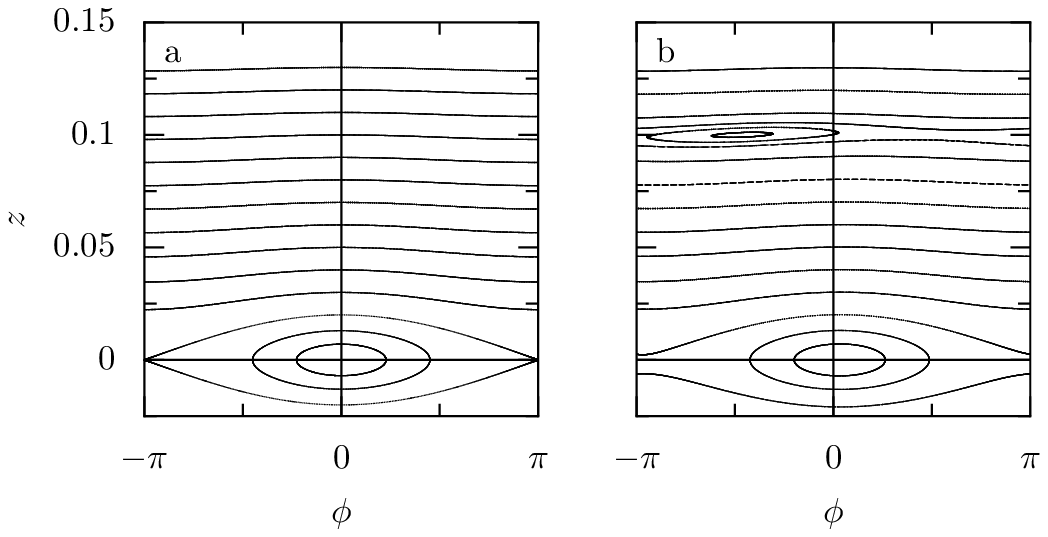}}
\caption{\label{fig:poincare} Phase space portrait of the undriven
[$\varepsilon=0$ (a)] and the driven [$\varepsilon=100$ (b)]
undamped momentum-shortened pendulum Hamiltonian \eref{H:pendulum}
at stroboscopic times $t\mathop{\rm mod} 2\pi k/\Omega=\pi/2$. The
interaction is $\Lambda=10^4$ and the driving frequency
$\Omega=1000$. }
\end{figure}%
In order to put the central discussion into the proper context, it
is convenient to review the dynamics of the relative phase after
two independently prepared condensates are connected
\cite{Zapata2003a}. Before the connection, the phase is completely
undefined, which means that the Fock state of fixed particle number
is in a coherent superposition of different phase states. Upon
connection, various mechanisms involving quasiparticle dynamics
intervene to destroy the coherence between the different phase
states. As a result, the reduced density matrix of macroscopic
phase-number system becomes quasi-diagonal in the phase
representation, which is to say that the phase is effectively
measured among a menu of values uniformly distributed between 0
and $2\pi$. After this quick definition of the phase, a
phase-number Gaussian wave packet forms that evolves
semiclassically in the parabolic tight-binding lattice formed by
the different number eigenstates. Due to the interaction term, the
system experiences a tilted lattice and, thus, undergoes Bloch
oscillations, which may be viewed as the ac Josephson response to
an approximately constant $\Delta \mu$ created by the number
imbalance with the possible concurrence of other factors. Its
trajectory follows one of the running solutions in phase space
diagram shown in \Fref{fig:poincare}a.  Thus, the system
oscillates around a nonzero number average $\bar z$ displaying the
so-called macroscopic quantum self-trapping (MQST)
\cite{Raghavan1999b}. However, these oscillations are not stable,
since the fluctuating condensate atom number in each well couples
to the many-quasiparticle field and experiences dissipation. Thus
the energy stored in the macroscopic degree of freedom decreases
and $\bar z$ decays slowly towards its equilibrium value
\cite{Zapata2003a}.

In the presence of an ac driving, the situation may change
qualitatively, since classically stable resonance islands form in
phase space for values of $z,\phi$ such that
\begin{equation} \label{averages}
\Lambda \bar{z} = k\Omega, \quad {\rmd\bar{z}}/{\rmd t}=0,
\end{equation}
where $k$ is an integer (see \Fref{fig:poincare}b). Here, the bars
indicate the time average over fast Bloch or external driving
oscillations, of frequencies $k\Omega$ and $\Omega$, respectively,
and the time derivative is meant over a longer time scale. In
general, the system displays running solutions weakly oscillating
around an average value $\bar z$ which slowly decays because of
dissipation. Such a decaying trajectory may or may not be trapped
in one of the attractive basins around the Shapiro resonances.

\begin{figure}[t]
\centerline{\includegraphics{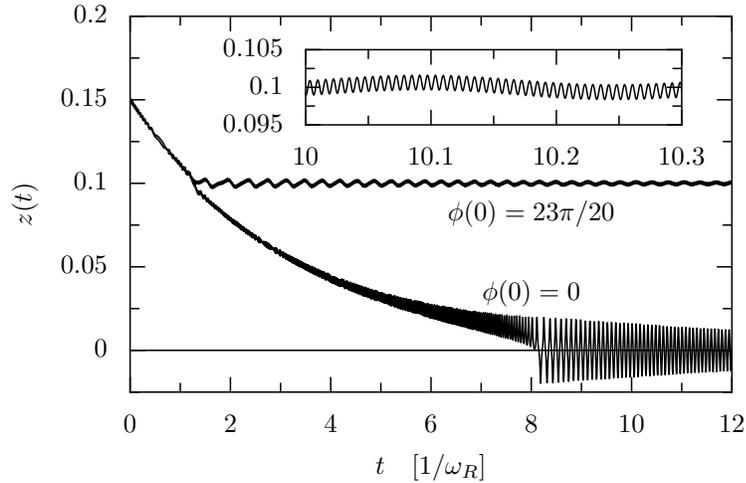}}
\caption{\label{fig:solutions} Population imbalance $z(t)$ for
interaction $\Lambda=10^4$, damping $\gamma=3\times 10^{-5}$,
driving amplitude $\varepsilon=100$, and frequency $\Omega=1000$
for two different initial values of the phase. The saturation
time-averaged value $\bar z = 0.1$ of the upper solution
corresponds to the chemical potential difference $\hbar\Omega$
($k=1$). The inset is a blow-up that resolves the fast underlying
Bloch (or MQST) oscillations. }
\end{figure}%

For a provisional answer to the central question formulated above,
we integrate numerically the equations of motion \eref{phi:dot}
and \eref{z:dot} in the Josephson regime ($\Lambda \gg 1$; we take
$\Lambda=10^4$), where the BEC Josephson junction displays, in the
absence of dissipation, stable running MQST solutions.  When
starting with an initial imbalance $z(0)$ that lies above a
resonance island, we find two qualitatively different types of
solutions: The first type (lower curve in \Fref{fig:solutions}) is
what one would typically expect, namely that, owing to the Ohmic
current, the imbalance decays until it leaves the MQST regime and
ultimately exhibits damped plasma oscillations. Except for the
small kink displayed when crossing the resonance zone, the
dynamics is qualitatively the same as in the undriven case.
However, for the same parameters but a different initial phase,
there exist also solutions with an intriguing behaviour (see upper
curve in \Fref{fig:solutions}): There, the system gets trapped in
the resonance island and the transient decay comes to a standstill
instead of continuing towards the true equilibrium solution
($z=0$). At long times, the number imbalance $z$ oscillates with
amplitude $\varepsilon$ around a non-zero value
$z_1=\Omega/\Lambda$.  We will show that then the average chemical
potential difference (in scaled units) is precisely a multiple of
the driving frequency. More specifically, we find
\begin{equation}
\label{mu:locked} \Delta\mu = k\Omega + \varepsilon\cos(\Omega t).
\end{equation}

We wish to remark that the BEC setup considered here differs
substantially form its superconducting analogs in that there the
stabilization of a voltage difference at the Shapiro resonance is
achieved with the concourse of an externally imposed current,
i.e., in an experiment a current source is needed.  Here by
contrast, the stabilization of a chemical-potential difference
cannot rely on such external current sources which for BECs are
not (yet) available.  As a drawback, the present type of
``voltage'' standard can be obtained only at the price that a
single run of the experiment does not warrant a stable solution.
In the following, we provide a detailed analysis of this type of
Shapiro resonances, estimating their parameter dependence and,
eventually, the probability that they can be spontaneously
realized.

\section{Realization of phase-locked solutions}
\label{sec:solutions}

Our central interest lies in the existence and the stability of
so-called phase-locked solutions, i.e., solutions $\phi(t)$ that
have only small fluctuations around a long-time average $\bar\phi
\approx k\Omega t$.  Then, since the main chemical-potential
difference between the two condensate fragments comes from the
interaction energy, the average number imbalance $\bar z$ settles
down at a value $z_k \equiv k\Omega/\Lambda$.  This motivates the
ansatz
\begin{eqnarray}
\phi &=& k\Omega t + \frac\varepsilon\Omega\sin(\Omega t)
+ \delta\phi , \label{ansatz:phi}
\\
z &=& \frac{k\Omega}{\Lambda} + \alpha\cos(k\Omega t) + \delta z ,
\label{ansatz:z}
\end{eqnarray}
where $\alpha$ denotes the yet unknown amplitude of the fast
residual Bloch oscillations that can be appreciated in the inset
of \Fref{fig:solutions}. We would like to choose $\alpha$ such
that the resulting $\delta \phi$ and $\delta z$ display only slow,
decaying oscillations around a stationary solution $\delta
\phi={\rm const.}$, $\delta {z}=0$ to which they tend at long
times. In other words, $\alpha$ should be such that, in the
absence of $\delta z$ and with $\delta \phi$ replaced by a
constant, equations \eref{ansatz:phi}, \eref{ansatz:z} correctly
capture the asymptotic system dynamics.

\begin{figure}[t]
\centerline{\includegraphics{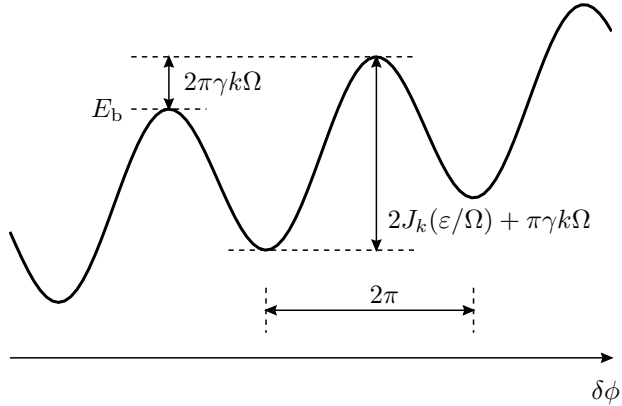}}
\caption{\label{fig:washboard} Sketch of the effective washboard
potential \eref{H:washboard} as a function of the phase difference
$\delta\phi$ for $\gamma k\Omega < J_k(\varepsilon/\Omega)$. }
\end{figure}

The equations of motion for $\delta\phi$ and $\delta z$ are
obtained by inserting 
\eref{ansatz:phi}, \eref{ansatz:z} into the original equations
\eref{phi:dot}, \eref{z:dot}. The resulting system of equations
involves rapidly oscillating coefficients of period $2\pi/\Omega$.
It can be shown that, if $\alpha=1/k\Omega$ and $\Omega \gg 1$,
then $\delta\phi$ and $\delta z$ vary more slowly than those
coefficients (a result which we have confirmed by numerical
studies), so that the different time-scales can be separated
consistently and all time-dependent coefficients can be replaced
by their {\it time-averages}. On the other hand, we are interested
in the case of two condensate fragments which are comparable in
size ($|z|\ll 1$). Thus we neglect within this analytic discussion
the momentum shortening and replace the square root in the
Hamiltonian \eref{H:pendulum} by unity \cite{Sols1999a}. Finally
we obtain the equations of motion
\begin{eqnarray}
\frac{\rmd}{\rmd t}\delta\phi &=& \Lambda \,\delta z ,
    \label{phi:dot:rwa}
\\
\frac{\rmd}{\rmd t}\delta\phi &=&
-J_k(\varepsilon/\Omega)\sin(\delta\phi)
    - \gamma k\Omega - \gamma\Lambda\,\delta z ,
    \label{z:dot:rwa}
\end{eqnarray}
which describe a dissipative particle in the static tilted washboard potential
sketched in \Fref{fig:washboard}.  They can be obtained from the Hamiltonian
\begin{equation}
\label{H:washboard} H(\delta\phi,\delta z)=\frac12 \Lambda\,\delta
z^2 -J_k(\varepsilon/\Omega)\cos(\delta\phi) + \gamma
k\Omega\,\delta\phi
\end{equation}
together with the dissipative force
$F_\mathrm{diss}=-\gamma\Lambda\,\delta z$.  The new canonical
coordinates $\delta z$ and $\delta\phi$ represent momentum and
position, respectively, and $J_k$ denotes the $k$-th Bessel
function of the first kind.  We emphasize two differences between
these equations and the undriven rigid pendulum Hamiltonian:
First, there is a tilt $\gamma k\Omega$ which originates from a
constant dissipative current caused by a finite number imbalance.
Despite its physical origin, it appears formally as a conservative
force. Second, the Josephson coupling energy for the phase shift
$\delta\phi$ is \textit{renormalized} by a factor
$J_k(\varepsilon/\Omega)$ which, for $\varepsilon\ll\Omega$,
renders it much smaller than the original Josephson energy.
Correspondingly, the plasma frequency is renormalized by a factor
$\sqrt{J_k(\varepsilon/\Omega)}$\footnote{If we had relaxed the
assumption $|z|\ll 1$, then $J_k(\varepsilon/\Omega)$ would have
appeared multiplied by a factor $\sqrt{1-\bar z ^2}$. However, one
must remember that, if $|z|$ were to become comparable to unity,
then the two-mode approximation might have to be revised.}. Once
$\delta\phi$ is trapped within one specific well, it exhibits
damped plasma oscillations with the renormalized frequency until
it ultimately comes to rest. In this stationary solution one has
$\delta\dot\phi= \delta \dot{z}=0$, which is possible thanks to
the cancellation of the first two terms in the r.h.s. of
\Eref{z:dot:rwa}. Therefore, one finds from \Eref{ansatz:phi} and
the Josephson relation $\dot\phi=\Delta\mu/\hbar$ the chemical
potential difference \eref{mu:locked}.

Having mapped the long-time dynamics of the originally time-dependent
situation to an
equivalent static problem, we are now able to derive the
conditions under which phase-locked solutions exist.  This results
in the restrictions of the parameters summarized in
\Tref{tab:restrictions}.  Obviously, the effective washboard
potential in the Hamiltonian \eref{H:washboard} possesses minima
only for $\gamma k\Omega < J_k(\varepsilon/\Omega)$ which amounts
for small driving amplitudes $\varepsilon\ll\Omega$ to $\gamma
k\Omega \ll (\varepsilon/2\Omega)^k$.  This is most easily
satisfied for $k=1$, where the condition for the existence of
wells reads $\varepsilon
> 2\gamma\Omega^2$.  Since on the other hand, the driving should
be weak, $\varepsilon\ll\Omega$, both conditions on the driving
amplitude $\varepsilon$ can be fulfilled simultaneously only for
sufficiently small dissipation, $\gamma\ll 1/\Omega$.

Also the interaction strength $\Lambda$ obeys restrictions, which
are obtained in the following way:  Physically, the idea behind
our scheme is to counterbalance a dissipative current with the
help of ac driving, thereby stabilizing a MQST solution against
friction. The condition for operating clearly in a regime where
MQST dominates is $2/\sqrt{\Lambda}\ll \bar z$
\cite{Raghavan1999b,Zapata2003a}. On the other hand, the two
condensate fragments should not differ too much in size, thus,
$\bar z\ll 1$. Since we aim at stabilizing the imbalance $\bar
z=\Omega/\Lambda$, fulfilling both conditions requires a
sufficiently large interaction strength, typically $\Lambda\gtrsim
10^3$.

A further physical reason for operating the Bose-Josephson
junction in the (interaction-dominated) Josephson limit stems from
an important property of phase-locked solutions, namely, that
generally the centre of an attractive basin is not truly at rest,
but follows the trajectory of a particle in the absence of driving
and dissipation \cite{Arnold1984a}.  In the Josephson regime, such
a behaviour corresponds to the MQST solutions exhibiting Bloch
oscillations with amplitude $\alpha=1/\Lambda\bar z$ around an
average value $\bar z$, as can be appreciated in \Eref{ansatz:z}.
At resonance, their relative amplitude becomes $\alpha/\bar
z=1/\Lambda\bar z^2=\Lambda/\Omega^2$. Ideally, for the
realization of stable phase-locked solutions, this amplitude
should be small, as is the case in the Josephson regime
($\Lambda\gg1$).  On the contrary in the (non-interacting) Rabi
regime ($\Lambda < 1$) \cite{Sols1999a}, the undamped and undriven
pendulum exhibits large number (Rabi) oscillations around $z=0$,
i.e.\ there is no MQST. This results in large oscillations of any
attractive basin which render BJJs in the Rabi regime useless for
the present purposes.

We conclude this section with a comparison between Shapiro
resonances and another well-known, apparently similar dynamical
phenomenon. It has been shown \cite{Holthaus2001a} that for a
high-frequency driving with $\Omega\gg\Delta\mu/\hbar$, the
Josephson energy acquires a factor $J_0(\varepsilon/\Omega)$ and
that, consequently, the coherent exchange of atoms is brought to a
standstill if $\varepsilon/\Omega$ is chosen as a zero of the
Bessel function $J_0$.  Such an effect amounts to an interacting
version of the so-called coherent destruction of tunnelling (CDT)
predicted for single particles in bistable potentials
\cite{Grossmann1991a}. We emphasize that the present phenomenon is
different in two respects: First, CDT takes place at zeros of
Bessel functions and, thus, requires large driving amplitudes,
while here, the driving amplitude is much lower.  Second, CDT
fades out under the influence of dissipation, while in our case,
moderate dissipation is essential for the convergence towards a
phase-locked solution.

\Table{\label{tab:restrictions} Conditions for the existence of
phase-locked solutions which are suitable for a chemical-potential
standard with $\Delta\mu=\hbar\Omega$ (i.e. $\bar z=\Omega/\Lambda$ and
$k=1$).}
\begin{tabular}{@{}ll}
\br
restriction & physical significance \\
\mr
$\varepsilon\ll\Omega$         & chemical potential difference much larger than its modulation\\
$\varepsilon>2\gamma\Omega^2$  & existence of stable wells in the washboard potential \\
$\Lambda\ll\Omega^2$        & operation within MQST regime, Bloch oscillations small \\
$\Lambda\gg\Omega$          & number imbalance small \\
\br
\end{tabular}
\endTable

\section{Locking probability}
\label{sec:probability}

After investigating which is the parameter regime that permits
stable phase-locked solutions, we turn to a remaining intriguing
question: What is the probability that such a solution is hit by
starting with a random initial phase? To give a crude estimate for
the answer, we employ again the analogy to the dissipative motion
in a tilted washboard potential.

Let us assume that we start out of resonance with an imbalance
$z_1<z<z_2$, with $z_k= k\Omega/\Lambda$, that corresponds to
$\delta z>0$, i.e., to a particle moving uphill in the tilted
washboard of \Fref{fig:washboard}. Having only finite energy, the
particle will bounce in one specific well.  At the entry point,
the energy will be in the range $[E_\mathrm{b} ,
E_\mathrm{b}+2\pi\gamma k\Omega]$, where $E_\mathrm{b}$ is the
potential energy at the top of the lower barrier.  While moving
within the well during one cycle, the particle dissipates the
energy
\begin{equation}
\label{E:diss}
E_\mathrm{diss}\approx -\int F_\mathrm{diss}\, \rmd\phi \approx
2\pi\gamma\Lambda\sqrt{2\varepsilon/\Lambda\Omega}  .
\end{equation}
Here, we have estimated the maximum value of the dissipative force
$F_\mathrm{diss}=-\gamma\Lambda\,\delta z$ [cf.\ \Eref{z:dot:rwa}]
from the maximum kinetic energy $\frac{1}{2}\Lambda\,\delta
z^2\approx 2J_k(\varepsilon/\Omega)\approx \varepsilon/\Omega$ for
$k=1$ and $\gamma\ll\varepsilon/\Omega^2$.
Assuming for the particle's phase space trajectory the shape of
an ellipse, yields the r.h.s.\ of \Eref{E:diss}.
If the initial energy minus the
dissipated energy lies below the barrier, the particle
will end up at rest in the well.  By assuming that the random
initial phases translate into equally distributed initial
energies, we find that this happens with probability
\begin{equation}
\label{probability} w = \frac{E_\mathrm{diss}}{2\pi\gamma\Omega}
\approx \sqrt{\frac{2\varepsilon\Lambda}{\Omega^3}} ,
\end{equation}
if $E_\mathrm{diss}< 2\pi\gamma\Omega$, and $w=1$ otherwise.
Ideally, one would like to have $w=1$ to ensure phase locking in
all runs. However, it is easy to show from the constraints in
\Tref{tab:restrictions} that this would require $\Omega \ll
\varepsilon$, which, for reasons already indicated, is not of
physical interest. Thus one must content oneself with identifying
a range of parameters that make the probability of landing in a
resonance non-negligible.

One might conceive situations where the control of the relative
particle number before the connection (as well as of the
interaction parameters) were so good that the system could be
forced to be within the attractive basin from the start. While
designing $z(0)$ to be very close to the resonance value $\bar z =
\Omega/\Lambda$ would surely increase the probabilities of
relaxing towards the Shapiro resonance, we wish to emphasize here
that success can never be fully guaranteed because of the {\it
intrinsically random nature} of the initial phase $\phi(0)$, as
can be clearly inferred from inspection of \Fref{fig:poincare}b.

\begin{figure}[t]
\centerline{\includegraphics{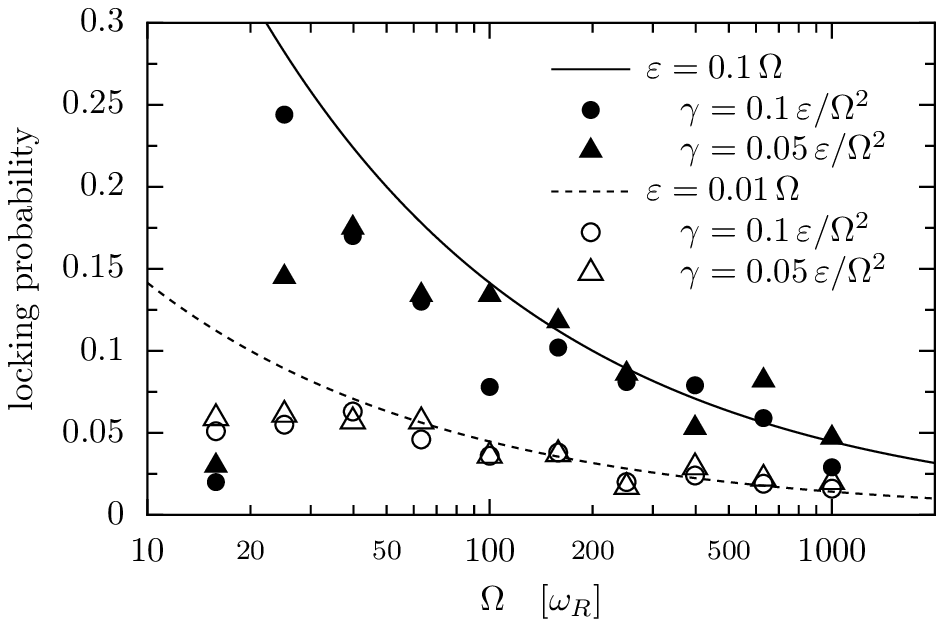}}
\caption{\label{fig:prob} Estimate \eref{probability} for the
locking probability compared to the fraction of phase-locked
solutions (symbols) obtained from numerical integrations of Eqs.\
\eref{phi:dot} and \eref{z:dot} with random initial phase.  The
driving amplitude is $\varepsilon=0.1\,\Omega$ (full line, filled
symbols) and $\varepsilon=0.01\,\Omega$ (broken line, open
symbols). The interaction is $\Lambda=\Omega/z_1$ where $z_1=0.1$,
and the dissipation $\gamma = 0.1\,\varepsilon/\Omega^2$ (circles)
and $\gamma = 0.05\,\varepsilon/\Omega^2$ (triangles).  Each data
point is obtained from 1000 simulation runs.}
\end{figure}%
To confirm the analytical estimate \eref{probability} for the
locking probability, we have integrated numerically the equations
of motion \eref{phi:dot} and \eref{z:dot} starting with a number
imbalance $z(0)=1.5\,\Omega/\Lambda$, and a random initial phase
$\phi(0)$. The parameters have always been chosen such that
$\varepsilon\ll\Omega\ll\Lambda$ and $\gamma\ll 1/\Omega$
according to the requirements derived in the last section.
\Fref{fig:prob} compares the analytical estimate
\eref{probability} to the fraction of numerical runs that converge
to a phase-locked solution.  The numerical and the analytical
result agree well in the regime $\Omega\gtrsim30$. For
$\Omega=100$, the locking probability typically assumes values of
order 10\%. We attribute the worse agreement between theory and
simulation for $\Omega \lesssim 30$ to the fact that, in such a
range, $\Omega$ becomes so small that the requirement $\Lambda \ll
\Omega^2$ (see \Tref{tab:restrictions}) can no longer be satisfied
if, as is the case in \Fref{fig:prob}, $\Lambda$ is constrained to
be $\Omega/z_1$ with $z_1=0.1$. An extreme case of disagreement is
found for the lowest frequency considered when $\varepsilon=0.1\,
\Omega$.

The Ohmic dissipation which we have assumed in our previous
analysis is only justified for temperatures well above the
chemical potential difference \cite{Zapata1998a}.  For a typical
condensate however, the chemical potential is of the same order as
the temperature and, thus, the dissipative current may not follow
a linear law.  Therefore, as a last item within the classical
analysis, we consider the more general dissipative current $\dot
z|_{\rm diss} = -g(z)$,
which might depend on various parameters like, e.g., the
interaction strength and the temperature. Using the ansatz
\eref{ansatz:z}, we linearize $g(z)\approx -g(z_k)-g'(z_k)\,\delta
z$ and repeat the analysis from above.  We find for $k=1$ again
the locking probability \eref{probability} but with an additional
factor $z_1g'(z_1)/g(z_1) = z_1[\ln g(z_1)]'$,
$z_1=\Omega/\Lambda$. On the other hand, the slope of the
washboard, i.e.\ the average dissipative current, acquires a
factor $g(z_1)$. For very low temperatures $k_BT\ll\Delta\mu$, for
instance, the quasiparticle-excitation decay mechanism discussed
in Ref.~\cite{Zapata2003a} yields $g(z)\propto z^2$, and therefore
the locking probability is enhanced by a factor of 2, but still is
of the same order.

\section{Quantum fluctuations}
\label{sec:quantum}

The classical treatment of the Bose-Josephson junction given above,
has its limitations due to quantum fluctuations of the number $n$ and
the phase $\phi$.  They come from the commutation relation
$[n,e^{i\phi}]=e^{i\phi}$ which amounts for small phase uncertainty to
\begin{equation}
\label{uncertainty}
[z,\phi]=-2i/N .
\end{equation}
This leads basically to two constraints for the washboard
potential: First, the potential wells of a washboard potential are
quantum mechanically metastable since a particle will tunnel out
at a rate \cite{Hanggi1990a}
\begin{equation}
\kappa = \omega_0 e^{-2\pi E_0/\hbar\omega_0}.
\end{equation}
To justify the classical treatment from above,
$1/\kappa$ must be larger than all other time scales of the
problem. A second, related point is that the phase-space region
corresponding to the metastable well must support sufficiently
many quantum states to treat both the number and the phase as
continuous classical variables.  For a well of depth $E_0$ with
curvature $\omega_0^2$, the number of quantum states can be
estimated as $m=E_0/\hbar\omega_0$. From both arguments, we
conclude that the ratio of the potential depth and the energy
quantum of a small oscillation at the bottom of the well
determines whether we operate in the classical limit $E_0 \gg
\hbar\omega_0$.

To determine the connection between $E_0$, $\hbar\omega_0$ and
the non-standard commutation relation \eref{uncertainty} and
our scaled parameters for $k=1$, we have to accomplish the
replacements
\begin{equation}
E_0 \to \frac{\varepsilon}{\Omega},\quad
\hbar \to \frac{2}{N},\quad
\omega_0^2 \to \frac{\varepsilon\Lambda}{2\Omega} .
\end{equation}
This results in
\begin{equation}
\frac{E_0}{\hbar\omega_0}=N\sqrt{\frac{\varepsilon}{2\Omega\Lambda}},
\end{equation}
which is essentially the ratio between the Josephson coupling
energy and the renormalized plasma frequency in the metastable
well. For typical parameters used above
($\varepsilon=0.01\,\Omega$, $\Lambda=10\,\Omega$, $\Omega \gtrsim
100$), a condensate consisting of $N\gtrsim 10^5$ atoms supports
$m\approx 100$ states and yields an escape rate that is
practically zero. Therefore, we do not expect any relevant quantum
correction.

\section{Conclusions}
\label{sec:conclusion}

We have investigated a realistic setup that may provide the basis
for a standard of chemical potential difference between weakly
connected atomic Bose-Einstein condensates. Due to the still
rudimentary development of atom circuit technology, we have
focused on schemes that do not require the coupling to a an
external circuit, thus staying away from straightforward analogues
of well-tested superconducting devices. In particular, we have
investigated the possibility of connecting two separate
condensates in the presence of ac driving in such a way that, as
the system relaxes towards equilibrium, it has an appreciable
probability of being trapped in a Shapiro resonance for which
$\Delta \mu$ is exactly an integer multiple of $hf$, with $f$ the
driving frequency. If the resolution of the imaging process is
good enough, a frequency $\Omega=2\pi f$ can be
found satisfying simultaneously
\begin{equation} \label{visibility}
\Lambda\, \Delta z < \Omega \ll \Lambda.
\end{equation}
While the second inequality ensures a small number imbalance (see
\Tref{tab:restrictions}), the first one expresses the experimental
ability to resolve two different Shapiro plateaus, $\Delta z <
z_{k+1}-z_k=\Omega/\Lambda$. Therefore, under any reasonable
visibility conditions, it should be possible to identify the
integer ratio $\Delta\mu/h f$, i.e., we can know the precise
Shapiro step where the double condensate has become locked. Thus,
we see that the main requirement of a precision measurement is
satisfied, namely, that, with the help of an accurately controlled
frequency, a poor measurement of a physical quantity (the atom
number imbalance) enables a fine measurement of another physical
quantity (the chemical potential difference).

Instead of being a hindrance, dissipation provides here a crucial
help, since it allows the system to spontaneously relax towards a
stable resonance where the chemical potential difference is
guaranteed to have a precise value. We have identified a number of
parameter constraints that must be satisfied in order to have an
optimum control of the Shapiro resonance. These include the need
to have a weak driving signal, the operation within the collective
Josephson regime, the need to keep average number imbalance small,
the requirement of even smaller Bloch oscillations, the necessity
to optimize the probability of landing in the desired Shapiro step
out of a random initial phase, and the robustness against escape
by quantum tunnelling from the metastable state.

Fortunately, this set of constraints does still allow for a window
of realistic parameters within which accurate chemical potential
differences could be realized.  Let us for instance consider
$N=10^6$ $^{23}$Na atoms in a split trap with frequency
$\omega_\mathrm{ho} = 2\pi\times 100\,\mathrm{Hz}$.  The bulk
properties of the condensate are readily estimated within the
Thomas-Fermi approximation to yield the chemical potential
$\mu\approx 70\,\hbar\omega_\mathrm{ho}$ and the mean field energy
$E_A(N)\approx\hbar N^{7/5}\times 0.01\,\mathrm{s}^{-1}$.
Estimating the Josephson coupling energy, respectively the Rabi
frequency, is more tedious and relies sensitively on the
assumptions made for the shape and the size of the barrier.  We
assume here a Gaussian shape of width $6\,\mathrm{\mu m}$ and
height of $1.05\,\mu$ such that for a driving amplitude
$\varepsilon=0.01\,\Omega$, the top of the barrier lies always
above the instantaneous chemical potential of both wells. Along
the lines of Refs.\ \cite{Zapata1998a,Sols1999a}, we obtain the
effective action $S_0=1.7\,\hbar$ and, thus, the Rabi frequency
$\omega_R\approx 2\pi\times 0.05\,\mathrm{Hz}$.  This finally
yields the effective interaction $\Lambda\approx 10^4$ that we
have assumed above.  The (bare) Josephson plasma frequency is
approximately $\omega_\mathrm{JP}=2\pi\times 5\,\mathrm{Hz}$; for
the resonant solution in \Fref{fig:solutions}, phase locking sets
in after $5\,\mathrm{s}$.

One might wonder what may be the use of a ``voltage" standard in
the absence of an atom circuit. First of all, no fundamental
reason prevents the use of the present scheme in future atom
circuits. Then, this concept will have to be tested against ideas
more directly borrowed from the technology of superconductors.
Second, already within the currently limited possibilities of
condensate transport, the realization of a Shapiro resonance may
provide a convenient playground for the investigation of novel BEC
scenarios. For instance, it seems possible to extend the
experimental work of Ref. \cite{Cataliotti2001a} to include an
external ac driving that locks a pair of wells into resonance. The
static behaviour of that pair of condensates should contrast with
the dynamic behaviour of atoms in neighbouring wells. That would
provide a most direct test of the Josephson effect in atomic
condensates. Third, a fine control of the chemical potential may
open the possibility of detailed checks of our current
understanding of the novel many-body problem posed by trapped
quantum gases. Quite likely, the effect of interactions will be
easily isolated from that of gravity or other well-controlled
external fields. Finally, there is the hope that the present work
may provide the basis for more efficient or practical concepts
that will make the emergent ``atomtronics" a precise technology.

\ack

This work has been supported by the Direcci\'on General de
Investigaci\'on Cient\'{\i}fica y T\'ecnica under Grant No.
BFM2001-0172, and by the Ram\'on Areces Foundation. We acknowledge
travel support from the European Science Foundation under the
program BEC 2000+.


\end{document}